\documentclass[pre,aps,twocolumn,amsmath,amssymb,showpacs,showkeys,floatfix]{revtex4-1}
\usepackage{graphicx}
\usepackage{amsfonts}
\usepackage{epsfig}
\usepackage{dcolumn}
\usepackage{amsthm}
\usepackage{color}

\usepackage{latexsym}
\usepackage{amscd}
\usepackage{hyperref}
\usepackage[mathscr]{eucal}

\usepackage{bm}      
\usepackage{ifpdf}
\usepackage{bbm}

\begin{document}

\newcommand{\newc}{\newcommand}

\newc{\beq}{\begin{equation}}
\newc{\eeq}{\begin{equation}}
\newc{\ovl}{\overline}
\newc{\bc}{\begin{center}}
\newc{\ec}{\end{center}}
\newc{\tr}{\mbox{tr}}
\newc{\pd}{\partial}
\newc{\dqv}{\delta\vec{q}}
\newc{\dpv}{\delta\vec{p}}
 \newc{\f}{\frac}

\title{Periodicity of quantum correlations in the quantum kicked top}
\author{Udaysinh T. Bhosale}
\email{udaybhosale0786@gmail.com}
\author{M. S. Santhanam}
\email{santh@iiserpune.ac.in}
\affiliation{Indian Institute of Science Education and Research, Dr. Homi Bhabha Road, Pune 411 008, India.}

\date{\today}

\begin{abstract}
Quantum kicked top is a fundamental model for time-dependent, chaotic Hamiltonian system and
has been realized in experiments as well.
As the quantum kicked top can be represented as a system of qubits, it is also popular as a
testbed for the study of measures of quantum correlations
such as entanglement, quantum discord and other multipartite entanglement measures. Further, earlier
studies on kicked top have led to a broad understanding of how these measures are affected by the
classical dynamical features.  In this work, relying on the invariance of quantum correlation measures 
under local unitary transformations, it is shown exactly these measures display periodic behaviour 
either as a function of time or as a function of the chaos parameter in this system.
As the kicked top has been experimentally realised using cold atoms as well as superconducting qubits,
it is pointed out that these periodicities must be factored in while choosing experimental
parameters so that repetitions can be avoided.

\end{abstract}

\maketitle

\section{Introduction}
\label{sec:Introduction}

Periodically kicked quantum systems are popular models of Hamiltonian
chaos. Their popularity, in part, arises from the relative ease of analysis. The quantum dynamics
of such systems can be reduced to a Floquet map,
while in the classical limit, the dynamics can be reduced to a set of
difference equations. The quantum kicked top is a prominent member of this class
and it physically represents a repeating sequence of free precession and state-dependent
rotation (kick). For sufficiently large kick strengths, the system displays
chaotic classical dynamics. Several approaches to experimental realization of quantum
kicked top were suggested \cite{Haake2000} and was attained using a cloud of cold Cs atoms
in the total hyperfine spin of its ground state interacting with
time-dependent magnetic fields \cite{Chaudhury09}.

In the last two decades, kicked top was widely used to study the interplay between chaotic 
dynamics and quantum correlations in the context of continued interest in quantum information and computation. 
The kicked top has a natural representation in terms of spins or qubits and this makes it a suitable choice for 
studies on entanglement. In this approach, the number of spins tending to infinity represents the classical limit of 
kicked top. Hence, this model continues to attract research interest 
\cite{Udaysinhbifurcation2017,Shohini2018,ArulChaos2018,ArulFewQubit2018,Arjendu2017,Quach2018} for the study of entanglement 
\cite{Lombardi2011,ShohiniGhose2008, Miller99,arul01,bandyo03,BandyoArulPRE} and its relation to classical 
dynamics \cite{Ghikas2007}, signatures of bifurcations on various quantum correlation measures 
\cite{Udaysinhbifurcation2017}, quantum-classical correspondence in the vicinity of periodic orbits 
\cite{Shohini2018} and quantum metrology \cite{Braun2018}. Measures of quantum 
correlations have been found to strongly correlate
with the qualitative nature of classical phase space, whether it is regular or chaotic 
\cite{Udaysinhbifurcation2017,Arjendu2017,VaibhavMadhok2015,BandyoArulPRE,Hiroshi2003,Lombardi2011}. 
In general, as demonstrated extensively in a series of papers using kicked tops 
\cite{Chaudhury09,Neill2016,Udaysinhbifurcation2017,Arjendu2017,VaibhavMadhok2015,BandyoArulPRE,Hiroshi2003,Lombardi2011}, the qualitative nature and details of classical dynamics influences 
entanglement. In addition, classical dynamical features such as the bifurcation also affect the
quantum correlation measures with interesting semiclassical consequences \cite{Udaysinhbifurcation2017}.
Similar results have been obtained for other measures of quantum correlations such
as quantum discord and Mayer-Wallach $Q$ measure. 

Unlike the earlier experimental effort \cite{Chaudhury09} involving manipulation of atomic and
nuclear spins, recently kicked top was realized in a system of just three
superconducting qubits (`spins') examining its behaviour in the deep quantum regime \cite{Neill2016}.
The latter experiment has verified the theoretically predicted connections \cite{Miller99,arul01,bandyo03}
between chaotic dynamics and bipartite entanglement. 
Quite remarkably, ergodic behaviour in this isolated quantum systems was demonstrated \cite{Neill2016}.
Surprisingly, a recent theoretical work has shown that even in the deep quantum limit possible 
with just two qubits, the system appears to take into account the nature of classical dynamics
in the vicinity of the phase space coordinates where the spin coherent state is
initially placed \cite{Arjendu2017}. Further, this work also hints that the entanglement entropy
might display (quasi-)periodic behaviour in time and also as a function of kick strength.
This observation, if generalized, has important implications for both experimental and theoretical work
on kicked tops. Let us consider a kicked top system with $j$ representing the total
spins and $k$ its kick strength. This corresponds to $2j$ number of spin-$1/2$ particles.
If a quantum correlation measure, say $A$, for this
kicked top displayed periodic behaviour, then for a given initial state we can expect the following functional
relations; $A(t;k,j)=A(t+T;k,j)$ or $A(t;k,j)=A(t;k+\kappa,j)$ representing periodic
behaviour in time $t$ and kick strength $k$ with periodicities, respectively, $T$ and $\kappa$.

This implies that for a fixed number of qubits quantum correlations will repeat
after a certain time period $T$ or after certain value kick strenth $\kappa$. Thus, generally
and crucially in an experimental context, the choice of $k$ and $j$ indirectly
sets the upper limit $T$ and $\kappa$ before repetitions begin to occur.
This argument can be turned around to derive another useful information. If an
experimental realization of the kicked top is
expected to maintain coherence for time-scale $\tau_{coh}$, then the question is about the
values of $k$ and $j$ that must be used in order to explore unique time evolution until time $\tau_{coh}$.
The mean coherence time $\tau_{coh}$ is generally a function of experimental (and
environmental) parameters, and together with values of $j$ and $k$ will uniquely determine
the relevant timescale for the experiment to be $\mbox{min}(\tau_{coh},T)$.
Thus, the present study of the periodicities in the kicked top will serve as a
crucial guide for experimental efforts to make the appropriate choice of parameters.

In this work, we show exactly that the time variation of quantum correlations of kicked top displays non-trivial 
periodicity provided the total spin $j=1$ and kick strength is of the form $k=r\pi/s$, $r$ and $s$ being integers. 
This includes the special case of two qubits, $j=1$, already reported in Ref.\cite{Arjendu2017}. Further, it is 
also shown that for any $j > 1$, though quantum correlations do not show temporal periodicity, they display 
periodic behaviour in kick strength $k$. Thus, this periodicity holds good in the
semiclassical limit of large $j$ as well. The structure of the paper is as follows: In Sec.~\ref{sec:Background} 
the measures of quantum correlations are introduced. In Sec.~\ref{sec:KickedTop} the kicked top model is 
introduced. In Sec.~\ref{sec:periodicity} analytical results on the periodicity of quantum correlations as a 
function of chaos parameter $k$ are given. In Sec.~\ref{sec:reflection} reflection symmetry of phase space in 
$k$ and its experimental consequences are discussed. In Sec.~\ref{sec:timeperiodicity} analytical results on 
time periodicity for the case of a two-qubit kicked top is studied.

\section{Measures of quantum correlations}
\label{sec:Background}
\subsection{von Neumann entropy}

Let us consider a standard bipartite system $A\otimes B$ composed of two smaller subsystems denoted as
$A$ and $B$, having Hilbert spaces $\mathcal{H}{_A}{^{(N)}}$ and $\mathcal{H}{_B}{^{(M)}}$ 
(with dimensions $N$ and $M$) respectively. For simplicity, $N\leq M$ can be assumed and
the full system belongs to the product Hilbert space 
$\mathcal{H}{_{AB}^{(MN)}}= \mathcal{H}{_A}{^{(N)}} \otimes \mathcal{H}{_B}{^{(M)}}$. 
Consider a normalized pure state
$|\psi\rangle=\sum_{i=1}^N \sum_{\alpha=1}^M c_{i,\alpha} |i\rangle\otimes|\alpha\rangle$ 
of the full system $A \otimes B$, where $|i\rangle\otimes|\alpha\rangle$ is the orthonormal 
basis of $\mathcal{H}{_{AB}}$. Its density matrix is $\rho=|\psi\rangle\langle\psi|$ 
satisfying the
Tr[$\rho$]=1 condition. The reduced density matrix of the subsystem $A$ is obtained by tracing out $B$ i.e.
$\rho_A= \mbox{Tr}_B[\rho]=\sum_{\alpha=1}^{M} \langle \alpha| \rho|\alpha \rangle$. Similarly, the subsystem $B$ is 
described by $\rho_B= \mbox{Tr}_A[\rho]$.
The singular value decomposition of the matrix $c_{i,\alpha}$ gives the following
Schmidt decomposition form:
\begin{equation}\label{Eq:SchmidtDecom}
 |\psi\rangle=\sum_{i=1}^{N}\sqrt{\lambda_i}~ | u_i^A\rangle \otimes | v_i^B \rangle
\end{equation}
where $|u_i^A\rangle$ and $|v_i^B \rangle$ are the eigenvectors of $\rho_A$ and $\rho_B$ respectively,
with the same eigenvalues $\lambda_i$. The eigenvalues $\lambda_i\in [0,1]$ are such that 
$\sum_{i=1}^N \lambda_i=1$. The remaining $M-N$ eigenvalues of $\rho_B$ are identically equal to zero.

Given the Schmidt eigenvalues $\lambda_i$ ($i=1\ldots N$), entanglement between $A$ and $B$, where 
von Neumann entropy is used as a measure, is given as follows:
\begin{equation}\label{Eq:vonNeumannEntropy}
S_{VN}=-\mbox{tr}(\rho_A\log\rho_A)=-\sum_{i=1}^{N}\lambda_i\ln(\lambda_i).
\end{equation} 
This is a good measure of entanglement for a bipartite pure state \cite{Bennett96,Zyczkowski06}. 
It satisfies $0\leq S_{VN}\leq \ln(N)$, where zero corresponds to a separable state and
$\ln(N)$ corresponds to a maximally entangled state.


\subsection{Quantum Discord}

Quantum discord measures all possible quantum correlations including and those beyond 
entanglement in a quantum state \cite{Ollivier,vedral}. 
This method involves removing the classical correlations from the total correlations of the system.
Now the procedure to evaluate discord will be given in detail \cite{Udaysinhbifurcation2017}.
For a bipartite quantum system having density matrix $\rho_{AB}$,
total correlations are quantified by the quantum mutual information given by,
\begin{eqnarray}
{\mathcal I}(B:A)  &=& {\mathcal H}(B) + {\mathcal H}(A) - {\mathcal H}(B,A) \label{eq5}.
\end{eqnarray}
On the other hand, the classical mutual information, based on Baye's rule, is given by
\begin{eqnarray}
I(B:A) &=& H(B) - H(B|A) , 
\label{eq6}
\end{eqnarray}
where $H(B)$ denotes the Shannon entropy of $B$. The conditional entropy $H(B|A)$ is defined 
as the average of the Shannon entropies of system $B$ conditioned on the values of $A$.
It can be thought of as the ignorance of $B$ given the information 
about $A$ \cite{nielsenbook}.

The quantum measurements on the subsystem $A$ are represented by a set of positive-operator valued measure
(POVM) $\{\Pi_{i}^{}\}$, such that the conditioned state of $B$ for given outcome $i$ is equal to
\begin{equation}
 \rho_{B|i}=\mbox{Tr}_{A}(\Pi_i \rho_{AB})/p_i\;\;\mbox{and}\;\; p_i=\mbox{Tr}_{A,B}(\Pi_i \rho_{AB})
\end{equation}
and its entropy is $\tilde{\mathcal H}_{\{\Pi_i\}} (B|A)=\sum_i p_i {\mathcal H}(\rho_{B|i})$.
In this case, the quantum mutual information is equal to 
${\mathcal J}_{\{\Pi_i\}} (B:A)= {\mathcal H}(B)- \tilde{\mathcal H}_{\{\Pi_i\}} (B|A) $.
Maximizing this over all possible measurement sets $\{\Pi_i\}$ one obtains 
\begin{eqnarray}
{\mathcal J}(B:A)&=&\mbox{max}_{\{\Pi_i\}}\left({\mathcal H}(B)- \tilde{\mathcal H}_{\{\Pi_i\}} (B|A)\right)\nonumber\\
&=&{\mathcal H}(B)-\tilde{\mathcal H}(B|A)
\end{eqnarray}
where $\tilde{\mathcal H}(B|A)=\mbox{min}_{\{\Pi_i\}}\tilde{\mathcal H}_{\{\Pi_i\}}(B|A)$. The minimum value is 
achieved using rank-one POVMs due to concave nature of the conditional entropy over the set of convex POVMs 
\cite{animesh_nullity}.
%
By taking $\{\Pi_i\}$ as rank-one POVMs, the quantum discord is defined as
${\mathcal D}(B:A)={\mathcal I}(B:A)-{\mathcal J}(B:A)$, such that
\begin{eqnarray}
{\mathcal D}(B:A) = {\mathcal H}(A)-{\mathcal H}(B,A)
+\mbox{min}_{\{\Pi_i\}} \tilde{\mathcal H}_{\{\Pi_i\}} (B|A).
\end{eqnarray}
The quantum discord is shown to be non-negative for all quantum states 
\cite{ZurekQuantumDiscord2000,Ollivier,animesh_nullity} and is subadditive \cite{VaibhavMadhok11}.
For the bipartite pure state, the quantum discord is shown to be equal to the von Neumann entropy \cite{Ollivier,vedral}.

\subsection{Concurrence and the 3-tangle}

Concurrence \cite{wootters98,Wootters01} is a measure of entanglement present between two qubits.
This measure was used to study phase transition in the Heisenberg chain \cite{Amico2008}.
Given two qubit density matrix $\rho_{AB}$, firstly the spin-flipped state 
$\tilde\rho_{AB}=\sigma_y\otimes\sigma_y\rho_{AB}^*\sigma_y\otimes\sigma_y$ is calculated, where 
$\sigma_y$ is the Pauli matrix and the complex conjugation is done in the standard basis.
Then the eigenvalues of the non-Hermitian matrix $\rho_{AB}\tilde{\rho}_{AB}$ are obtained,
which are all real and non-negative such that $\lambda_4\leq \lambda_3\leq \lambda_2\leq \lambda_1$. 
Then, the concurrence $C_{12}=C(\rho_{AB})$ is equal to 
\begin{equation}
\mbox{max}~(0,\sqrt{\lambda_1}-\sqrt{\lambda_2}- \sqrt{\lambda_3}- \sqrt{\lambda_4})
\end{equation}
and $ 0\le C_{12}\le 1$. It is zero for separable state and one for maximally entangled state. It is shown
that the entanglement of formation \cite{Bennett96a} of $\rho_{AB}$ is a monotonic function of concurrence 
\cite{Wooters,Wootersentform}. For the Bell state, concurrence is equal to one.

The 3-tangle is a pure multipartite entanglement measure for pure as well as mixed three-qubit states 
\cite{Coffman}. For the case of a three-qubit pure state, it is given by 
$\tau=C^2_{1(23)}-C^2_{12}-C^2_{13}$ \cite{Coffman}, where $C_{ij}$ measures the concurrence between
$i$-th and $j$-th qubits. The quantity $C_{1(23)}$ is the concurrence between qubit $1$ and the pair 
of qubits $2$ and $3$. 
This is because in a three-qubit pure state, the reduced density matrix of qubits $2$ and $3$ is of rank-$2$.
The 3-tangle $\tau$ is permutationally invariant and satisfies $0\le \tau \le 1$ \cite{Coffman}. 
For given concurrence $C_{12}$ the maximum 3-tangle $\tau$ a three-qubit pure state can have has been calculated
\cite{Udaysinh2016}. States satisfying these limits have also been evaluated.

\subsection{Meyer and Wallach $Q$ measure}
This multipartite entanglement measure \cite{Meyer02} was studied earlier in the context of spin Hamiltonians
\cite{ArulLakshminarayan2005,Karthik07,Brown08}, system of spin-bosons \cite{Lambert2005}
and how it is affected due to the classical bifurcation in the kicked top model \cite{Udaysinhbifurcation2017}.
The geometric multipartite entanglement measure $Q$ is shown to be related to one-qubit 
purities \cite{Brennen}, making its calculation and interpretation straightforward.
If $\rho_i$ is the reduced density matrix of the $i$th spin obtained by tracing out the rest of the spins in a 
$N$ qubit pure state then the $Q$ measure is defined as follows:
 \begin{equation}
\label{Eq:Qmeasure}
 Q(\psi)=2 \left( 1-\frac{1}{N}\sum_{i=1}^{N}\mbox{Tr}(\rho_i^2) \right).
 \end{equation}
The relation in Eq.~(\ref{Eq:Qmeasure}) between $Q$ and the single spin reduced density matrix purities has 
led to a generalization of  $Q$
measure to multiqudit states as well as for various other bipartite splits \cite{Scott2004}.

\section{kicked top}
\label{sec:KickedTop}
The quantum kicked top is characterized by an angular momentum vector ${\bf J} = ( J_x, J_y, J_z )$ and
its components obey the standard algebra of angular momentum. Here, the Planck's constant has been set to unity.
The Hamiltonian governing the dynamics of the top is given by
\begin{equation}
H(t) = p J_y + \frac{k}{2j} J_{z}^{2} \sum_{n=-\infty}^{+\infty} \delta (t-n).
\label{eq:single}
\end{equation}
The first term represents the free precession of the top around $y-$axis with angular
frequency $p$ while the second term is periodic $\delta$-kicks applied to the top.
Each kick gives a torsion about the $z-$axis by an angle $(k/2j) ~J_z$.
Here, $k$ is called as the chaos parameter or the kick strength.
For  $k=0$ the classical limit of Eq.~(\ref{eq:single}) is integrable and 
for $k > 0$ it becomes increasingly chaotic.
The corresponding period-one Floquet operator of the Hamiltonian in Eq.~(\ref{eq:single}) is given as follows:
\begin{equation}
U=\exp\left(-i\frac{k}{2j} J_{z}^{2}\right) \exp\left(-i p J_{y}\right). 
\label{FloquetOperator}
\end{equation}
The Hilbert space dimension is equal to $2j+1$ implies that the dynamics can be explored without any
truncation of the Hilbert space.
The kicked top has been realized in various experimental test beds, in hyperfine levels of cold Cs atoms
and coupled superconducting qubits \cite{Chaudhury09,Neill2016}, in which $p=\pi/2$. 
In \cite{Neill2016}, it was found that the time-averaged
von Neumann entropy showed the clear resemblance with the corresponding
classical phase-space.

The quantum kicked top for given angular momentum $j$ can be considered equivalent to a quantum 
simulation of a collection of $N=2j$ number of qubits (spin-half particles)
whose evolution is restricted to the subspace which is symmetric under the exchange of the qubits.
The state vector is restricted to a symmetric subspace spanned by
the basis states $\{|j,m\rangle ; (m = -j,-j + 1, . . . ,j )\}$ where $j=N/2$.
The basis states satisfy the property 
$S_z|j,m\rangle=m|j,m\rangle$ and $S_{\pm}|j,m\rangle=\sqrt{(j\mp m)(j\pm m+1)} |j,m \pm 1\rangle$ where $S_z$ and 
$S_{\pm}$ are collective spin operators \cite{DavidBook,HuMing2008}. The states 
$\{|j,m\rangle\}$ are also known as Dicke states. Thus, it is a multiqubit system 
whose collective behaviour is governed by the Hamiltonian in Eq.~(\ref{eq:single}) 
and the quantum correlations between any two qubits can be studied.

The classical phase space is displayed in Fig.~\ref{fig:phasespaceplot1} as a function of coordinates $\theta$ and 
$\phi$. In order to explore quantum dynamics in the kicked top, spin-coherent states
\cite{Hakke87,Arecchi1972,Glauber1976,PuriBook} pointing along the direction of
$\theta_0$ and $\phi_0$ are constructed and are evolved under the action of the Floquet operator.
The classical map for the kicked top is given as follows \cite{Haakebook,Hakke87}:
\begin{subequations}
\begin{eqnarray}
X^{\prime} &=& (X \cos p + Z \sin p) \cos\left(k\left(Z\cos p - X \sin p\right)\right)\nonumber\\ 
&& -Y \sin\left(k\left(Z \cos p- X \sin p\right)\right),\\
Y^{\prime} &=& (X \cos p + Z \sin p)\sin \left(k\left(Z \cos p-X \sin p\right)\right) \nonumber\\ 
&& +Y \cos\left(k\left(Z \cos p-X\sin p\right)\right),\\
Z^{\prime} &=& -X \sin p + Z \cos p.
\end{eqnarray} 
\label{eq:ClassicalMap}
\end{subequations}
Here, the dynamical variables $(X,Y,Z)$ satisfy the constraint $X^2+Y^2+Z^2=1$, i.e., they are
restricted to be on the unit sphere. Thus, it is possible to parameterize them
in terms of the polar angle $\theta$ and the azimuthal angle $\phi$ as 
$X = \sin \theta \cos \phi$, $Y = \sin \theta \sin \phi$ and $Z = \cos \theta$. 
First, the map in Eq.~(\ref{eq:ClassicalMap}) is evolved and then the values of 
$(\theta,\phi)$ are determined using the inverse relations, which are not shown here.

Another feature of this map is that under the transformation $k\rightarrow-k$ the phase-space 
is reflected about $\theta=\pi/2$. This is because $k \rightarrow -k $ is equivalent to the 
transformation $X \rightarrow -X $ and $Z \rightarrow -Z $ in Eq.(\ref{eq:ClassicalMap}). 
This implies $Z' \rightarrow -Z' $ which results in  $\theta \rightarrow \pi-\theta$. Thus, 
the phase-space corresponding to $k$ and $-k$ are isomorphic to each other. This can be seen 
from Figs.~\ref{fig:phasespaceplot1}(b) and \ref{fig:phasespaceplot3}(c), as well as from 
Figs.~\ref{fig:phasespaceplot1}(c) and \ref{fig:phasespaceplot3}(d). This has experimental implications which 
will be discussed in later part of the paper.

\begin{figure}[t]
\begin{center}
\includegraphics*[scale=0.325]{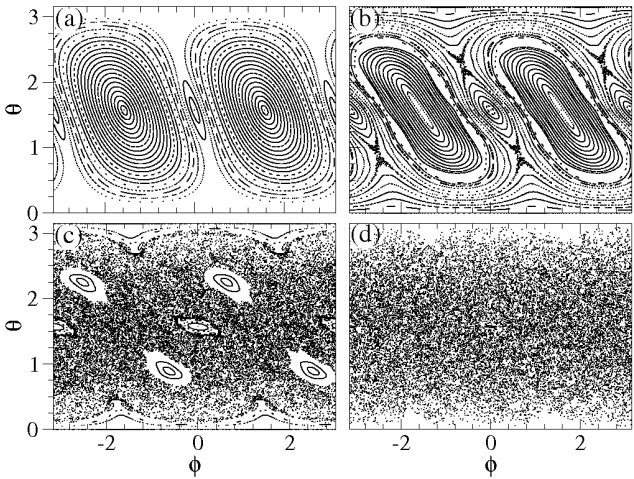}
\caption{(Color online) Phase-space pictures of the classical kicked top for $p=\pi/2$ and
(a) $k=1$, (b) $k=2$, (c) $k=3$ and (d) $k=6$.}
\label{fig:phasespaceplot1}
\end{center}
\end{figure}

\begin{figure}[t]
\begin{center}
\includegraphics*[scale=0.325]{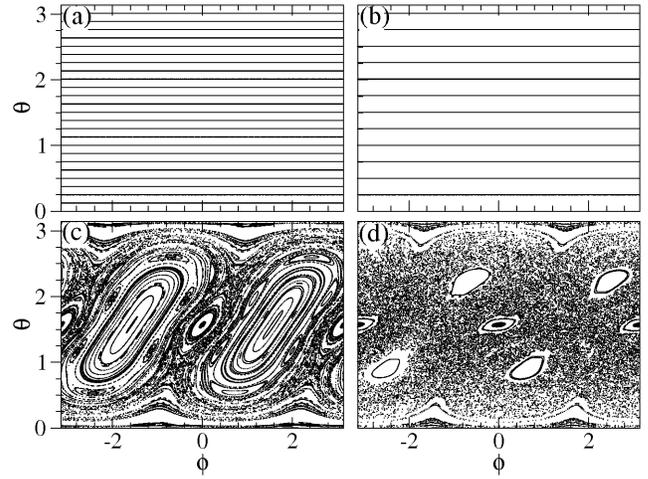}
\caption{(Color online) Phase-space pictures of the classical kicked top for $k=3\pi/5$ and
(a) $p=\pi$ and (b) $p=2\pi$. Same for $p=\pi/2$ and (c) $k=-2$ and (d) $k=-3$.}
\label{fig:phasespaceplot3}
\end{center}
\end{figure}

\subsubsection{Classical map for various values of $p$}
In this work, the model is studied for various values of $p$. Thus, it will be helpful to study the corresponding
map equations and the phase-space. First, the case of $p=\pi/2$ is considered. 
In this case, due to additional symmetries, a simpler classical map can be obtained
and was studied in detail in
Refs.~\cite{Hakke87,BandyoArulPRE,VaibhavMadhok2015,Udaysinhbifurcation2017,Shohini2018,Chaudhury09,Neill2016}.
In this case, the map given in Eq.~(\ref{eq:ClassicalMap}) reduces to
\begin{eqnarray}
\label{eq:ClassicalMap1}
\begin{split}
X^{\prime} &= Z \cos\left(k X\right)+Y \sin\left(k X\right),\\ 
Y^{\prime} &=  Y\cos\left(k X\right)-Z \sin\left(k X\right),\\ 
Z^{\prime} &= -X.
\end{split}
\end{eqnarray} 
The phase-space obtained using these equations is displyed in Fig.~\ref{fig:phasespaceplot1}.
It can be seen that for $k=1$ and $k=2$ the phase-space is mostly covered by regular orbits.
The trivial fixed points at $(\theta,\phi)=(\pi/2,\pm\pi/2)$ can be seen in Fig.~\ref{fig:phasespaceplot1}(a)
and Fig.~\ref{fig:phasespaceplot1}(b) becomes unstable at $k=2$. As $k$ is increased further 
the chaotic regions are increased. At $k=6$ the phase-space is covered mostly by the chaotic sea 
with very tiny regular islands.
 
The map for $p=3\pi/2$ can be obtained from that of $p=\pi/2$ by the transformation 
$X^{\prime}\rightarrow -X^{\prime}$ and $Z^{\prime}\rightarrow -Z^{\prime}$. This implies
$\phi\rightarrow-\phi$ and  $\theta \rightarrow \pi-\theta$ which are reflections about $\phi=0$ and 
$\theta=\pi/2$. 
Thus, the phase-space, as well as other properties, can be obtained by taking these reflections.
 
Now consider the case of $p=\pi$. In this case using Eq.~(\ref{eq:ClassicalMap}) the
classical map is obtained as follows:
\begin{eqnarray}
\label{eq:ClassicalMap2}
\begin{split}
X^{\prime} &=  Y \sin\left(k Z\right) - X \cos\left(k Z\right),\\ 
Y^{\prime} &=  Y\cos\left(k Z\right)- X \sin\left(k Z\right),\\ 
Z^{\prime} &= -Z.
\end{split}
\end{eqnarray} 
The phase-space is plotted in Fig.~\ref{fig:phasespaceplot3}(a). It can be seen that there is no 
fully developed chaos since for given initial $Z$ the angle $\theta$ oscillates between $\cos^{-1} Z$
and $\pi - \cos^{-1} Z$. Both these values are reflection about $\pi/2$ which can also be seen in the figure.

For the case $p=2\pi$ the map equations are
\begin{eqnarray}
\label{eq:ClassicalMap3}
\begin{split}
X^{\prime} &=   X \cos\left(k Z\right)- Y \sin\left(k Z\right),\\ 
Y^{\prime} &=   X \sin\left(k Z\right) + Y\cos\left(k Z\right), \\ 
Z^{\prime} &= Z.
\end{split}
\end{eqnarray} 
The phase-space is plotted in Fig.~\ref{fig:phasespaceplot3}(b). In this case too there is no fully developed 
chaos and for given initial $Z$ the angle $\theta$ remains fixed at $\cos^{-1} Z$.

\section{Periodicity of quantum correlations as a function of chaos parameter}
\label{sec:periodicity}
In this section, it will be shown analytically and through numerical 
simulations that the quantum correlations display periodicity
as a function of kick strength $k$. In particular, it will be shown that for 
a fixed value of $j$ and for a given initial state, the quantum correlations 
are periodic in $k$, with $\kappa=2j\pi$ being its periodicity.
\setcounter{subsubsection}{0}
\subsubsection{$j=1$ case}

Let us consider the simplest case of $j=1$ which is equivalent to two qubits.
Then, the basis states are $|1,-1\rangle$, $|1,0\rangle$ and $|1,1\rangle$.
The standard two qubit basis states are $\{|0\rangle_1|0\rangle_2$, $|0\rangle_1|1\rangle_2$,
$|1\rangle_1|0\rangle_2$, $|1\rangle_1|1\rangle_2\}$ (subscripts label qubits) such 
that $\sigma_z |0\rangle = -|0\rangle$ and
$\sigma_z |1\rangle= |1\rangle $. Both the basis states are related to each other by
$|1,-1\rangle=|0\rangle_1|0\rangle_2$, $|1,1\rangle=|1\rangle_1|1\rangle_2$ and 
$|1,0\rangle=(|0\rangle_1|1\rangle_2 + |1\rangle_1|0\rangle_2)/\sqrt{2}$.

Setting $j=1$ in Eq. \ref{FloquetOperator}, the corresponding Floquet operator is 
\begin{equation}
U=\exp\left(-i\frac{k}{2} J_{z}^{2}\right) \exp\left(-i p J_{y}\right). 
\label{Jis1FloquetOperator}
\end{equation}
It can be seen 
that when $k\rightarrow k+2\pi$ one obtains 
\begin{equation}
U \rightarrow \widehat{O}\, U ~~\mbox{where}~~ \widehat{O}=\exp\left(-i\pi J_{z}^{2}\right). 
\end{equation}
Thus, $U |\psi_j\rangle \rightarrow \widehat{O}\, U |\psi_j\rangle$
where $|\psi_j\rangle$ is any vector in the $|j,m\rangle$ basis. For $j=1$ case,
denoting the vector $U |\psi_1\rangle=[a,b,c]^T$.
Operator $\widehat{O}$ is diagonal in $\{|j=1,m\rangle\}$ basis i.e. $\widehat{O}=\mbox{diag}[-1,1,-1]$.
However, in the standard two-qubit basis it becomes
\begin{equation}
\widehat{O} = \left( \begin{matrix}
          -1 & 0 & 0 & 0\\
          0 & 1/2 & 1/2 & 0\\
          0 & 1/2 & 1/2 & 0\\         
          0 & 0 & 0  & -1\\
\end{matrix}\right).
\end{equation}
Thus, it can be seen that even though
$\widehat{O}$ is unitary in $\{|j=1,m\rangle\}$ basis, it is not so in the standard two-qubit basis.
{This implies that $\widehat{O}$ is not a local unitary but it will seen now that its action on any state in 
$\{|j=1,m\rangle\}$ basis does not change the quantum correlations among the qubits.}
%
Thus, in $\{|j,m\rangle\}$ basis $[a,b,c]^T \rightarrow \widehat{O}[a,b,c]^T=[-a,b,-c]^T$. It can be shown 
easily that in the 
standard two qubit basis states, $[a,b,c]^T$ becomes $|\chi_1\rangle=[a,b/\sqrt{2},b/\sqrt{2},c]^T$
whereas $[-a,b,-c]^T$ becomes $|\chi_1\rangle'=[-a,b/\sqrt{2},b/\sqrt{2},-c]^T$.
Thus, we have,
\begin{align}
|\chi_1\rangle = & a|1\rangle_1|1\rangle_2+(b/\sqrt{2}) \left(|1\rangle_1|0\rangle_2+|0\rangle_1|1\rangle_2 \right)+ \nonumber \\
               & c|0\rangle_1|0\rangle_2 \,\,\,\,\, \,\,\, \mbox{and} \nonumber \\
|\chi_1\rangle' = & -a|1\rangle_1|1\rangle_2+(b/\sqrt{2}) \left(|1\rangle_1|0\rangle_2+|0\rangle_1|1\rangle_2 \right)-  \nonumber \\
                & c|0\rangle_1|0\rangle_2.
\end{align}
It is seen that $|\chi_1\rangle$ and $|\chi_1\rangle'$ are related to each other by a
local unitary transformation, i.e.,
$|\chi_1\rangle'= -\sigma_z\otimes\sigma_z |\chi_1\rangle$.
Quantum correlation measures by definition are invariant under local unitary operations 
\cite{Horodeckirpm}. Using concurrence for two-qubit pure state \cite{Wootters01} it can be seen
to be equal to $2|b^2/2-ac|$ for both the states. These imply that the correlations are invariant
under the transformation $k\rightarrow k+2\pi$. This can be seen in Fig.~\ref{fig:jis1Entropy} where
von Neumann entropy shows a periodicity of $2\pi$ as a function of chaos parameter $k$.
 
\begin{figure}[t!]
\begin{center}
\includegraphics*[scale=0.80]{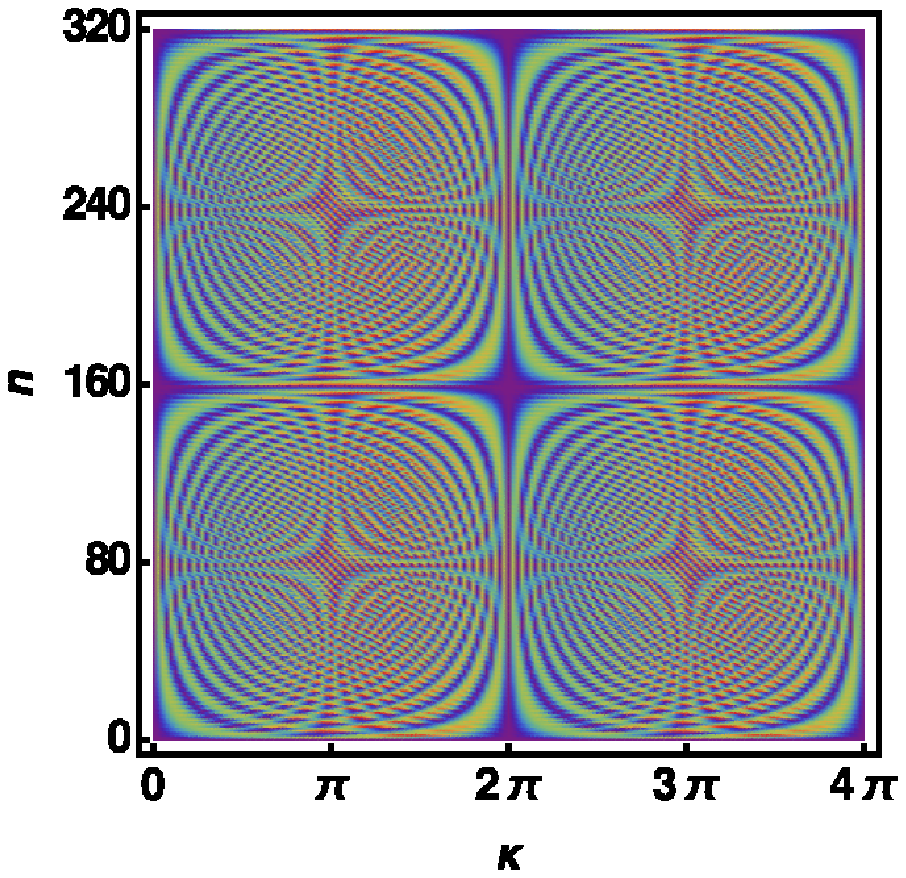}
\includegraphics*[scale=0.80]{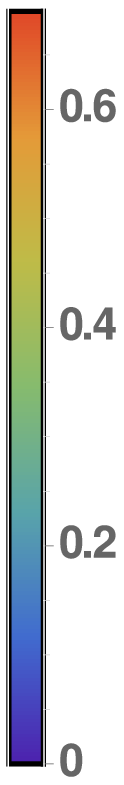}
\caption{(Color online) The von Neumann entropy ($S_{VN}$) is displayed for a two-qubit ($j=1$) kicked top
with parameters $p=\pi/2$ and $k=r\pi/40$ ($r=0\ldots 160$). The color bar by the side represents the
color code for the von Neumann entropy values.
Note the $2\pi$ periodicity in kick strength $k$ as well as the periodicity in time evident in this picture.}
\label{fig:jis1Entropy}
\end{center}
\end{figure}

\subsubsection{General $j$ case}
Let us consider the case of general $j$, beginning with even integer value for $j$.  
Here, the corresponding operator 
$\widehat{O}=\exp\left(-i\pi J_{z}^{2}\right)$ is diagonal matrix of order $2j+1$ in $\{|j,m\rangle\}$ basis,
i.e., $\widehat{O}=\mbox{diag}[1,-1,\ldots,-1,1]$.
The transformation $k\rightarrow k+2j\pi$ gives $U \rightarrow \widehat{O}\, U$.
The operator $\widehat{O}$ is diagonal matrix of dimension $2j+1$ in $\{|j,m\rangle\}$ basis i.e. 
$\widehat{O}=\mbox{diag}[1,-1,\ldots,-1,1]$. Now, the basis 
$\{|j,m\rangle\}$ will be written in the standard basis of qubits. For given value of $m$ there are 
$2j \choose j+m$ basis states superposed equally to form $|j,m\rangle$ where each of the basis state is 
such that $j+m/2$ qubits are in up-state $|1\rangle$ and remaining $j-m/2$ qubits are in down-state $|0\rangle$. 
In this paper, such a basis state will be called as $m-$particle state since it is an eigenvector of the total spin
operator $S_z$ with eigenvalue $m$. 
Thus, there are ${2j \choose j+m}$ $m-$particle states and the normalization constant after superposing
all such $m-$particle states is $1/\sqrt{{2j \choose j+m}}$. For example, 
$|j,1\rangle = (|1\rangle_1|0\rangle_2\ldots|0\rangle_{2j} +  |0\rangle_1|1\rangle_2\ldots|0\rangle_{2j}+ 
\ldots +|0\rangle_1|0\rangle_2  \ldots|1\rangle_{2j} )/\sqrt{{2j \choose 1}}$.

It is easily evident that $\widehat{O}$ is a block-diagonal matrix in $\{|j,m\rangle\}$ basis and can be
denoted as $\mbox{diag}[\widehat{O}_0,\widehat{O}_1,\ldots,\widehat{O}_{2j}]$. Similar to the $j=1$ case, 
 $\widehat{O}$ is unitary in $\{|j,m\rangle\}$ basis but it is no longer unitary when written 
in the standard $2j+1$ qubit basis. 
{Thus, $\widehat{O}$ is not a local unitary. But, we will now show that the quantum correlations 
remains invariant after $\widehat{O}$ acts on any state in the $\{|j,m\rangle\}$ basis. 
}
Here, each $\widehat{O}_n$ ($n=0,1,\dots 2j$) is a square matrix
of dimension $2j \choose n$ and each element in it is equal to $\exp\left(-i\pi n^{2}\right)/{2j \choose n}$, where
$n=j+m$ takes values in the range $0 \dots 2j$. It should be noted that each $\widehat{O}_n$ is written in the set of all
$n-$particle states. The vector $U|\psi_j\rangle$, in the $\{|j,m\rangle\}$ basis, is denoted as
$[c_0,c_1,c_2,\dots,c_{2j-1},c_{2j}]^T$. The same vector in the $m-$particle basis, $m=-j$ to $j$, becomes
$|\chi_j\rangle=[c_0',c_1',c_1',\dots,c_{2j-1}',c_{2j-1}',c_{2j}']^T$. In this, $c_n'= c_n/\sqrt{2j \choose j+m}$
and each $c_n'$ occurs $2j \choose n$ times in a sequence.
Thus, $\widehat{O}|\chi_j\rangle=\mbox{diag}[\widehat{O}_0,\widehat{O}_1,\ldots,\widehat{O}_{2j}]
[c_0',c_1',c_1',\dots,c_{2j}']^T$.

\begin{figure}[t!]
\begin{center}
\includegraphics*[scale=0.80]{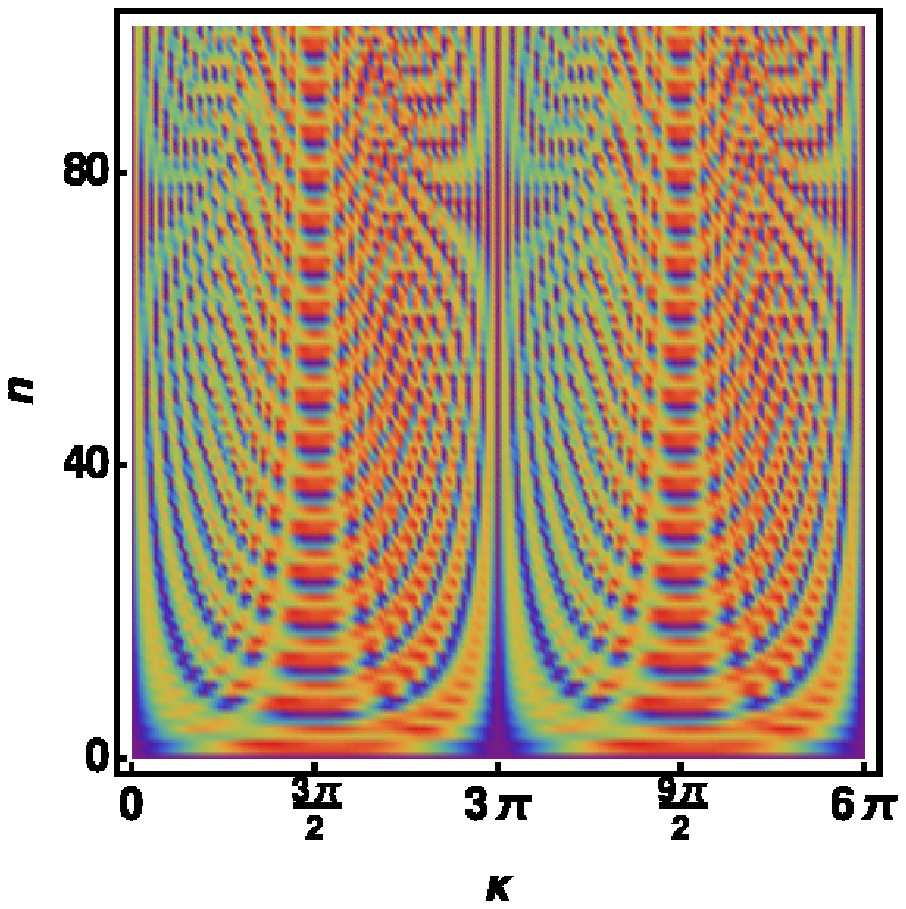}
\includegraphics*[scale=0.80]{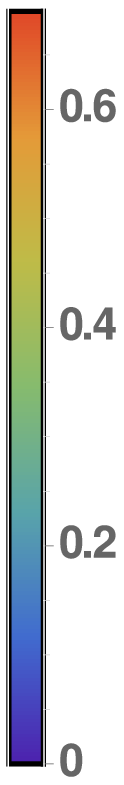}
\includegraphics*[scale=0.80]{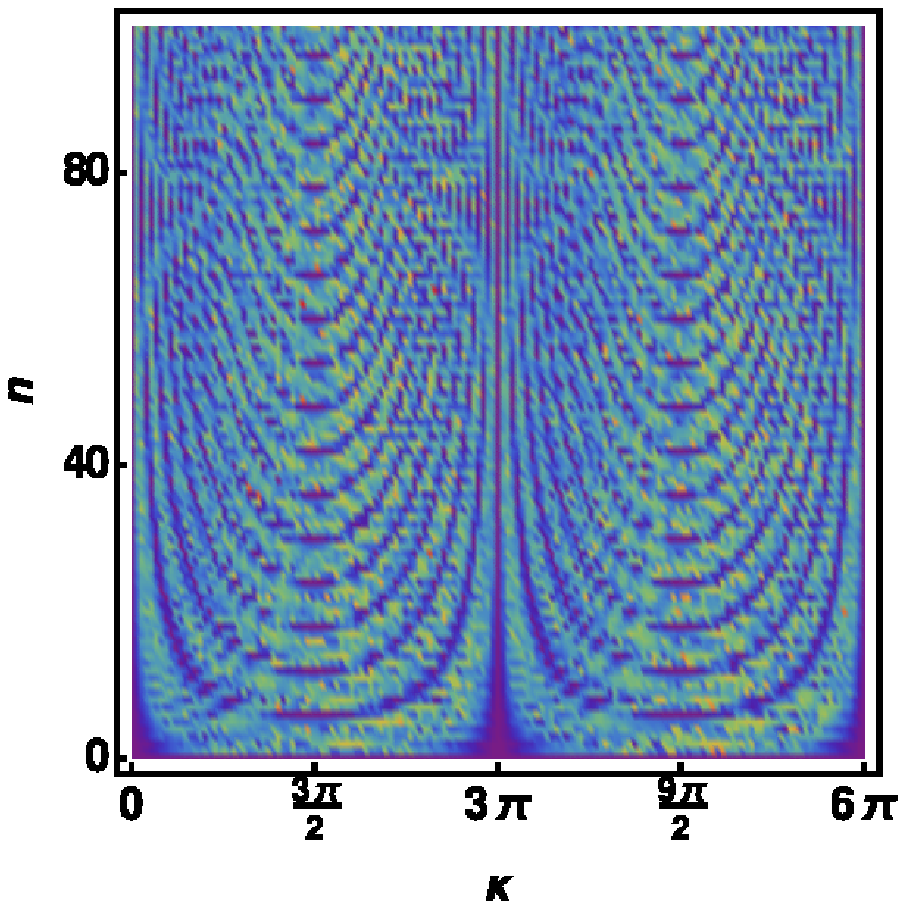}
\includegraphics*[scale=0.80]{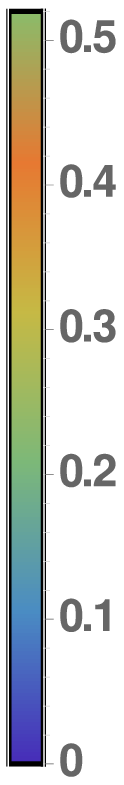}
\caption{(Color online) (top) von Neumann entropy ($S_{VN}$) of kicked top which is partitioned
as a single qubit and two qubits, (bottom) quantum discord $({\mathcal D})$ between any two qubits.
Both are plotted as function of kick strenght $k$ and time. In this, $j=3/2$. The values of
von Neumann entropy and discord are color coded using the color map shown by the side.}
\label{fig:jis3by2Q1}
\end{center}
\end{figure}

\begin{figure}[h!]
\begin{center}
\includegraphics*[scale=0.80]{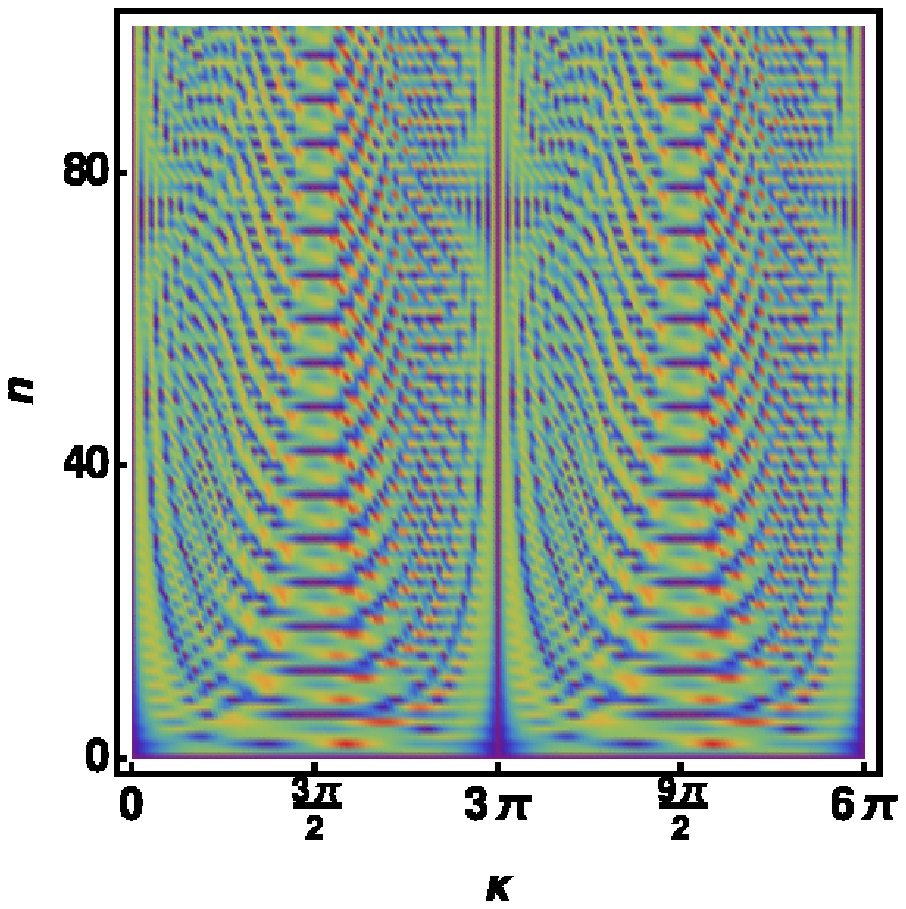}
\includegraphics*[scale=0.80]{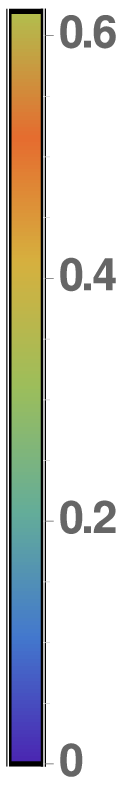}
\includegraphics*[scale=0.80]{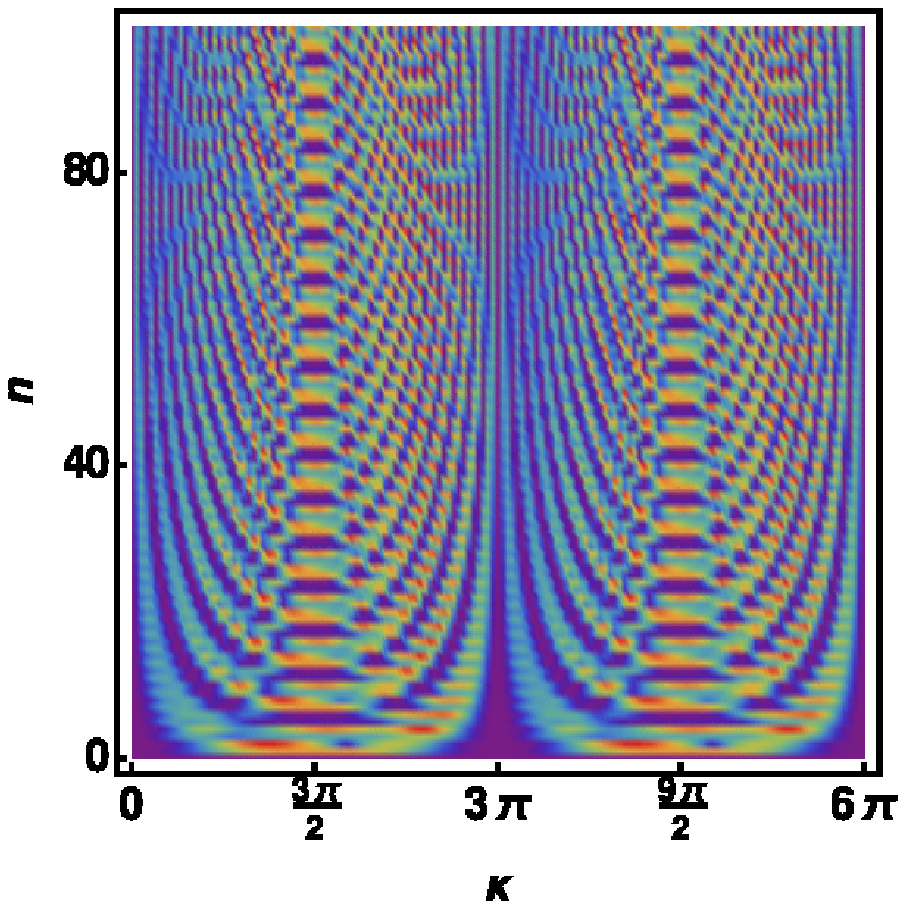}
\includegraphics*[scale=0.80]{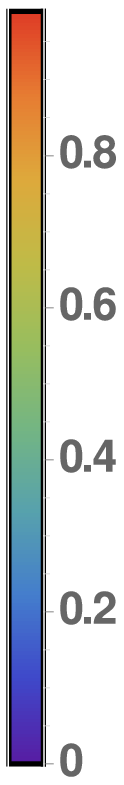}
\caption{(Color online) (top) Concurrence ($C_{12}$) between any two qubits,
(bottom) 3-tangle ($\tau$).
Both are plotted as function of kick strenght $k$ and time. In this, $j=3/2$. The concurrence
and 3-tangle values are color coded using the color map shown by the side.}
\label{fig:jis3by2Q2}
\end{center}
\end{figure}
 
Thus, it is seen that the matrix $\widehat{O}_0$ having dimension one gets multiplied by the column 
vector of dimention one containing $c_0'$,
the matrix $\widehat{O}_1$ having dimention $2j \choose 1$ 
gets multiplied by the column vector of dimention $2j \choose 1$ having $c_1'$ as its element at all the rows
and so on. Thus, in general the matrix $\widehat{O}_n$ of order $2j \choose n$ 
gets multiplied by the column vector of length
$2j \choose n$ having $c_n'$ as its element at all the rows. 

\begin{figure}[t!]
\begin{center}
\includegraphics*[scale=0.80]{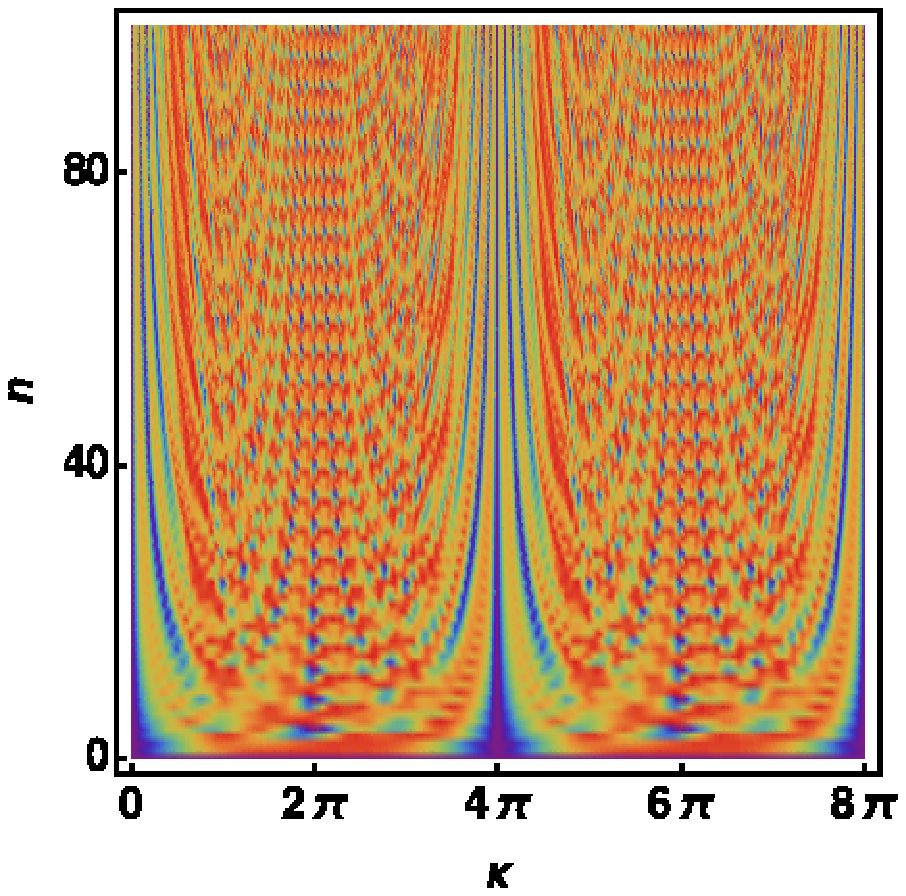}
\includegraphics*[scale=0.80]{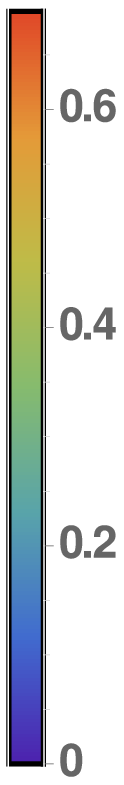}
\includegraphics*[scale=0.80]{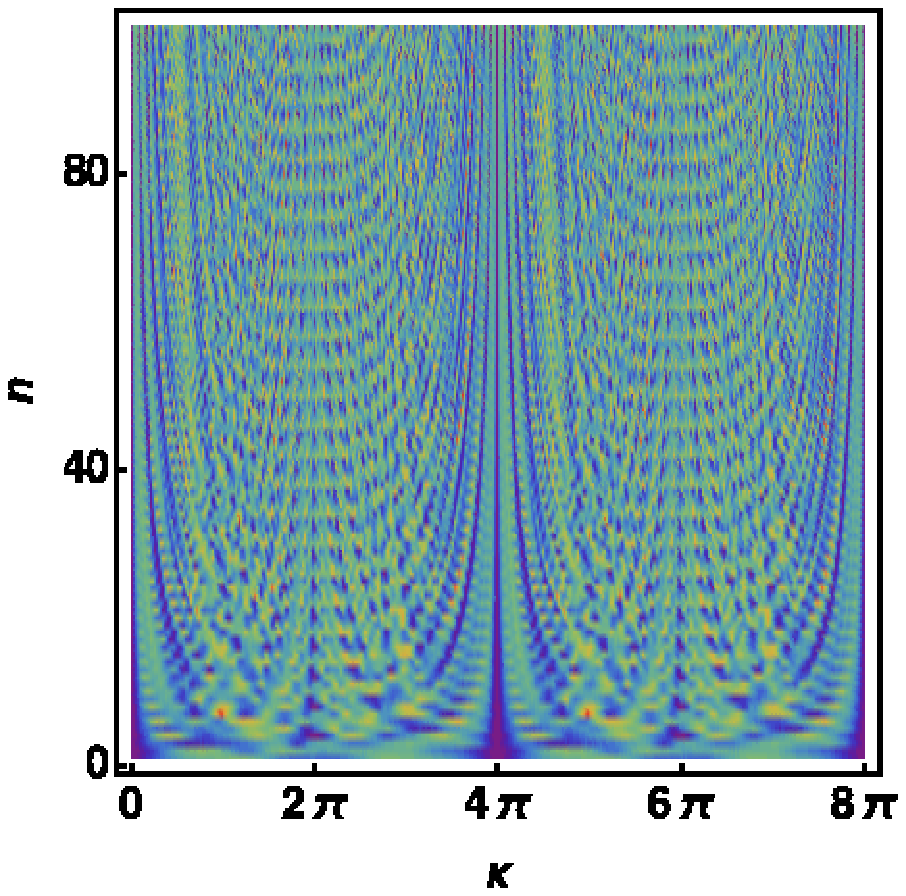}
\includegraphics*[scale=0.80]{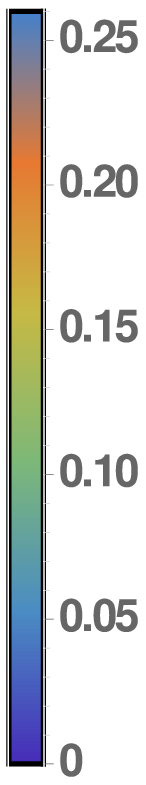}
\caption{(Color online) Same as Fig. \ref{fig:jis3by2Q1} for $j=2$.}
\label{fig:jis3Q1}
\end{center}
\end{figure}

\begin{figure}[t!]
\begin{center}
\includegraphics*[scale=0.80]{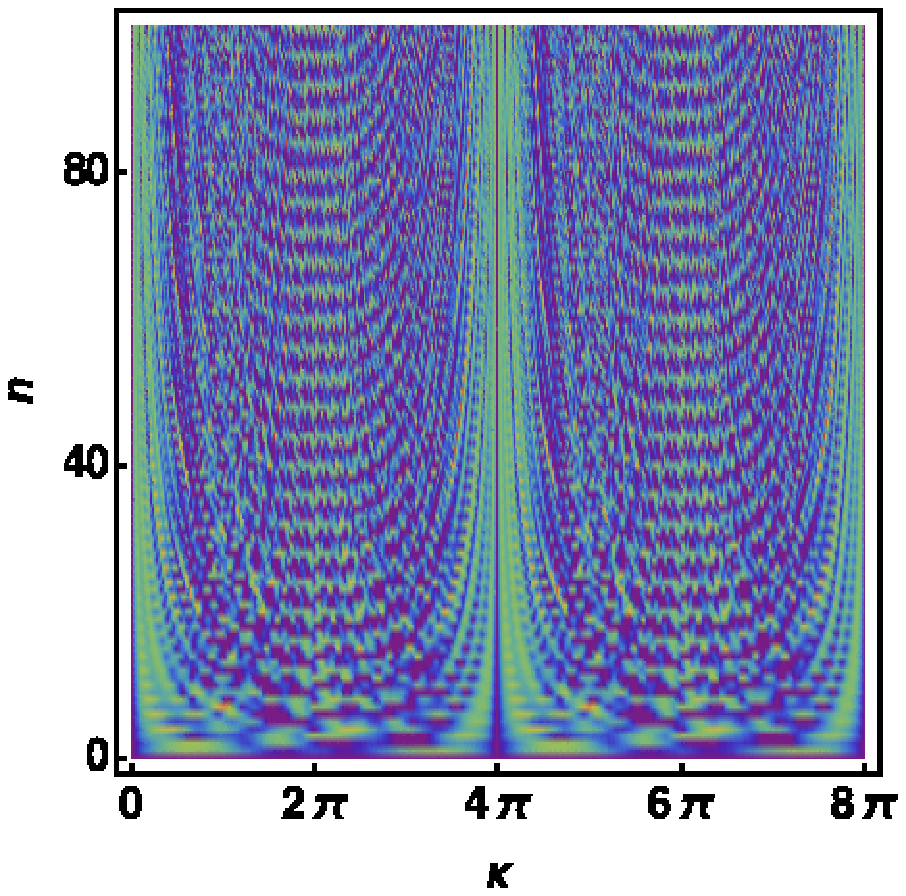}
\includegraphics*[scale=0.80]{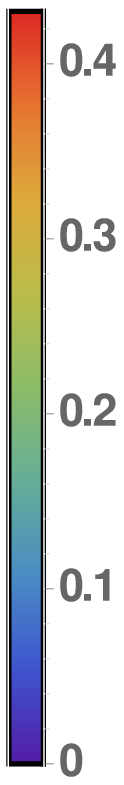}
\includegraphics*[scale=0.80]{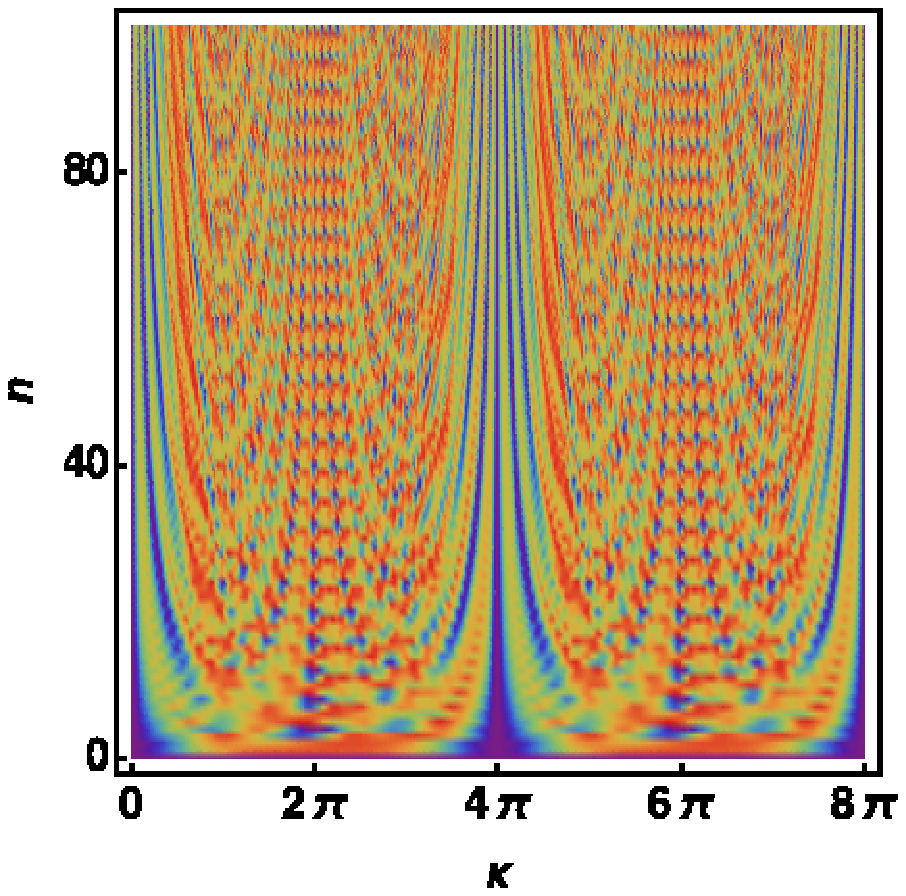}
\includegraphics*[scale=0.80]{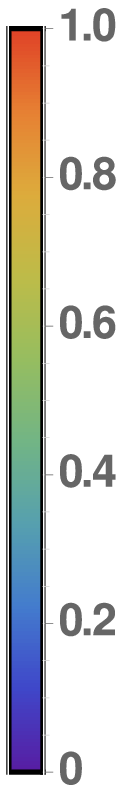}
\caption{(Color online)
(top) Concurrence ($C_{12}$) between any two qubits and (bottom) Meyer and Wallach $Q$ measure for $j=2$.
It is color coded using the color map shown by the side.}
\label{fig:jis3Q2}
\end{center}
\end{figure}

Let us denote this (unnormalized) column vector by $|\xi_j^n\rangle =[c_n',c_n',\ldots,c_n']^T $.
As pointed out earlier, $\widehat{O}_n$ is square matrix of order $2j \choose n$
with matrix elements $\exp\left(-i\pi n^{2}\right)/{2j \choose n}$.
This leads to
\begin{align}
\widehat{O}_n |\xi_j^n\rangle & = \exp\left(-i\pi n^{2}\right)[c_n',c_n',\ldots,c_n']^T. 
\end{align}
Thus, the final product becomes
\begin{align}
\widehat{O}|\chi_j\rangle=[c_0',-c_1',-c_1',\ldots,-c_{2j-1}',-c_{2j-1}',c_{2j}']^T.
\end{align}
When tranformed to $\{|j,m\rangle\}$ basis, it becomes $[c_0',-c_1',c_2',\ldots,-c_{2j-1}',c_{2j}']^T$.
It can also be written as $\sum_{n=0}^{2j} (-1)^n  c_n' |j,j-n\rangle$. Here $j$ is even and
using the properties of $|j,j-n\rangle$ it becomes
\begin{equation}
\left(\prod_{i=1}^{2j}\otimes \sigma_z^i \right) |j,j-n\rangle= (-1)^n |j,j-n\rangle,
\end{equation}
where the superscript denotes the qubit position.
Thus,
\begin{eqnarray*}
&&\sum_{n=0}^{2j} (-1)^n  c_n' |j,j-n\rangle=\left(\prod_{i=1}^{2j}\otimes \sigma_z^i  \right)\sum_{n=0}^{2j} c_n' |j,j-n\rangle\\
&=&\left(\prod_{i=1}^{2j}\otimes \sigma_z^i  \right)  [c_0',c_1',c_1',\dots,c_{2j-1}',c_{2j-1}',c_{2j}']^T\\
&=&\left(\prod_{i=1}^{2j}\otimes \sigma_z^i  \right) |\chi_j\rangle.
\end{eqnarray*}
Hence, 
\begin{equation}
\label{eq:ocap1}
\widehat{O}|\chi_j\rangle = \left(\prod_{i=1}^{2j}\otimes \sigma_z^i  \right) |\chi_j\rangle
\end{equation}
which implies
\begin{equation}
\label{eq:ocap2}
\widehat{O}U|\psi_j\rangle = \left(\prod_{i=1}^{2j}\otimes \sigma_z^i\right) U|\psi_j\rangle.
\end{equation}

Clearly, for the case of even $j$ as well, the two states are related to each other by local unitary operations.
Relying on the invariance of the quantum correlation measures under local unitary
operations \cite{Horodeckirpm}, which in this context implies invariance under $k\rightarrow k+2j\pi$,
it is inferred that the quantum correlations are periodic as a function of $k$ with period $2j\pi$. It must emphasized that the quantum correlations are periodic in $k$ even
for large value $j$, i.e, in the semiclassical limit as well. 
Similar result can be proved for the case of odd and half-integer values of $j$. 
This can be seen in the simulation results displayed in Figs.~\ref{fig:jis3by2Q1}, \ref{fig:jis3by2Q2}, 
\ref{fig:jis3Q1} and \ref{fig:jis3Q2}, where various quantum correlations show periodicity of $2j\pi$ as a 
function of chaos parameter $k$. Here, the initial coherent state is positioned at $\theta=2.5$ and $\phi=1.1$ 
for all values of $k$. It should be emphasized here that this result is valid only for any initial state
$|\psi_j\rangle$ in the symmetric subspace spanned by the basis states $\{|j,m\rangle\}$ which may or may not 
be an eigenstate of $J_z$. 
%
It should also be noticed from Eq.~(\ref{eq:ocap1}) that the operator
$\widehat{O}$ is non-unitary in the qubit basis while
$\prod_{i=1}^{2j}\otimes \sigma_z^i$ a local unitary operator in the same basis.
However, the result of their actions on the state $|\psi_j\rangle$ are equal.
\begin{figure}[t!]
\begin{center}
\includegraphics*[scale=0.29]{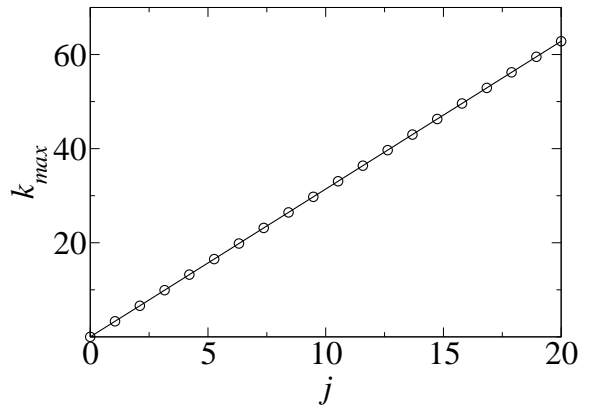}
\caption{(Color online) Maximum value of chaos parameter $k_{max}=j\pi$ such that phase space effects on quantum
correlations is unique as a function of number of qubits $j$.}
\label{fig:kmaximum}
\end{center}
\end{figure}

\section{Reflection symmetry in $k$ and experimental consequences}
\label{sec:reflection}

Now, consider two different values of chaos parameters $k_1$ and $k_2$ such that $0\leq k_1 \leq j\pi$ and 
$ j\pi \leq k_2 \leq 2 j\pi$. Further, they are related by $k_2=2j\pi-k_1$ representing a reflection symmetry about $j\pi$.
As the quantum correlations are periodic in $k$ with a period of $2j\pi$,
the time evolution of quantum correlations at $k=k_2$ is identical to that at
$k=-k_1$. As mentioned in Sec.~\ref{sec:KickedTop}, the phase space for $k$ and $-k$ are 
isomorphic to each other and are related by the transformation
$\theta\rightarrow\pi-\theta$. This implies that if an initial state is evolved for $k=k_2$ then it
is equivalent to the evolution of initial state for $k_1=2j\pi-k_2$ provided the initial positions of both the 
coherent states are related by  $\theta\rightarrow\pi-\theta$. We will call this a signature of phase space.

Thus, the combination of $2j\pi$ periodicity and symmetry in $k$ results in quantum correlations
that are symmetric about $k=j\pi$. In other words, for fixed value of  $j$, the maximum value 
of chaos parameter $k_{\mbox{max}}$ for which the phase space effects are unique is $j\pi$. 
Beyond $k=k_{\mbox{max}}$, the observed structure repeats itself. The maximum chaos parameter 
$k=k_{\mbox{max}}$ for the given number of qubits in the top is shown in Fig.~\ref{fig:kmaximum}. 
This result has implications for kicked top experiments. If two qubits are used to represent the kicked top, 
i.e. $j=1$, then one can observe the unique signatures of the phase space only up to $k=\pi$.
If three qubits are used, as done in the case of a recent experimental realization
reported in Ref. \cite{Neill2016}, one can observe the unique signatures of the phase space 
only upto $k=3\pi/2 \approx 4.71$ and so on.

\section{Time periodicity of quantum correlations for $j=1$}
\label{sec:timeperiodicity}

It can be seen from Fig.~\ref{fig:jis1Entropy} that the von Neumann entropy also exhibits periodicity in time 
for certain values of $k$. A similar effect, quasi-periodicity of entanglement, was also observed in 
Refs. \cite{ShohiniGhose2008,Arjendu2017}. 
The quantum discord between any two qubits was numerically shown to display quasi-periodic
modulations for initial states localized in the regular regions \cite{VaibhavMadhok2015}.
It was also pointed out that all the quantum expectation values are quasi-periodic in
in time due to the discreteness of the spectrum of Floquet operator \cite{Haakebook}

In this section, the $j=1$ case is considered and it is shown analytically that when $k$ is a rational 
multiple of $\pi$, and $p$ takes value from the set $\{0, \pi/2,\pi, 3\pi/2,2\pi \}$, the quantum correlations 
show periodic nature. We note that in the experiments reported in Refs. \cite{Chaudhury09,Neill2016},
$p=\pi/2$ is used. In Fig.~\ref{fig:jis1Entropy} the von Neumann 
entropy is plotted for $p=\pi/2$ and $k=r\pi/40$ such that $r=0$, $1$, $\ldots$, $160$. This gives the time 
period as $160$. This section is devoted to explaining this observation. Starting from 
Eq.~(\ref{Jis1FloquetOperator}) the matrix elements of the corresponding Floquet operator can be determined and 
assembled in matrix form.

\subsection{Case of $p=\pi/2$}
If $p=\pi/2$, then the Floquet operator reduces to
\begin{equation}
\label{}
U= \left( \begin{matrix}
\dfrac{e^{-ik/2}}{2} & \dfrac{-e^{-ik/2}}{\sqrt{2}} &  \dfrac{e^{-ik/2}}{2}\\
1/\sqrt{2} &               0  &   -1/\sqrt{2}\\
\dfrac{e^{-ik/2}}{2}  & \dfrac{ e^{-ik/2}}{\sqrt{2}} & \dfrac{e^{-ik/2}}{2}\\
\end{matrix}\right).
\end{equation}
Its eigenvalues are $\{ e^{-ik/2}, -i \, e^{-ik/4}, i \, e^{-ik/4} \}$ and the corresponding
eigenvectors are $[1/\sqrt{2},0,1/\sqrt{2}]^T$, $[-1/2,-i \, e^{ik/4}/\sqrt{2},1/2]^T$ and $[-1/2,i \, e^{ik/4}/\sqrt{2},1/2]^T$
respectively. Using these the Floquet operator for $n$th time can be obtained which is given as follows:

\begin{widetext}
\begin{equation}
\label{Eq:UNpiby2} U^n =
\frac{1}{4} \left( \begin{matrix}
2 \,e^{-ikn/2} + \left(-i e^{-ik/4} \right)^n   + \left(i e^{-ik/4} \right)^n  
& 
\dfrac{\left(-i e^{-ik/4} \right)^n   - \left(i e^{-ik/4} \right)^n }{i \, 2^{-1/2} \, e^{ik/4}}
&   2 \,e^{-ikn/2} - \left(-i e^{-ik/4} \right)^n   - \left(i e^{-ik/4} \right)^n  \\
\dfrac{\left(-i e^{-ik/4} \right)^n   - \left(i e^{-ik/4} \right)^n }{-i \, 2^{-1/2} \, e^{-ik/4}}
& 2 \left(  \left(-i e^{-ik/4} \right)^n   - \left(i e^{-ik/4} \right)^n  \right) 
&  \dfrac{\left(-i e^{-ik/4} \right)^n   - \left(i e^{-ik/4} \right)^n }{i \, 2^{-1/2} \, e^{-ik/4}}\\
2 \,e^{-ikn/2} - \left(-i e^{-ik/4} \right)^n   - \left(i e^{-ik/4} \right)^n 
& \dfrac{\left(-i e^{-ik/4} \right)^n   - \left(i e^{-ik/4} \right)^n }{-i \, 2^{-1/2} \, e^{ik/4}}
& 2 \,e^{-ikn/2} + \left(-i e^{-ik/4} \right)^n   + \left(i e^{-ik/4} \right)^n\\                
\end{matrix}\right).
\end{equation}
\end{widetext}
Now, we will consider the case of $k=r\pi/s$, for various choices of integral values of $r$ and $s$.
It will be proved that if $r$ is odd then the time period of quantum correlations is $T=4s$,
otherwise it is $T=2s$.

{\it Odd $r$ :} If $r$ is odd integer and time $n=4s$, the Eq. \ref{Eq:UNpiby2} simplifies to
 \begin{equation}
\label{} U^{4s} =
 \left( \begin{matrix}
0 & 0   &  1\\
0 & -1  &  0\\
1 & 0 & 0\\                
\end{matrix}\right).
\end{equation}
Thus, $U^{4s} [a,b,c]^T = [c,-b,a]^T$. In the two-qubit basis, this becomes
\begin{equation}
c|1\rangle_1|1\rangle_2-(b/\sqrt{2})(|1\rangle_1|0\rangle_2+|0\rangle_1|1\rangle_2)+
a|0\rangle_1|0\rangle_2.
\end{equation}
Now, this can be rewritten in the following form;
\begin{eqnarray}
\begin{split}
&(\sigma_z\otimes\sigma_z)(\sigma_x\otimes\sigma_x)(a|1\rangle_1|1\rangle_2+\\
&(b/\sqrt{2})(|1\rangle_1|0\rangle_2+|0\rangle_1|1\rangle_2)+c|0\rangle_1|0\rangle_2).
\end{split}
\end{eqnarray}
Hence, $[c,-b,a]^T = (\sigma_z\otimes\sigma_z)(\sigma_x\otimes\sigma_x) [a,b,c]^T $
implying that the two states are related to each other by local unitary transformation
supporting the claim for the periodicity of quantum correlations.

{\it Even $r$ :} In the case of even $r$, using Eq.~(\ref{Eq:UNpiby2}), one obtains
\begin{equation}
\label{}U^{2s} =
 \left( \begin{matrix}
\dfrac{1-(-1)^{r/2}}{2} & 0   &  \dfrac{1+(-1)^{r/2}}{2}\\
0 & -(-1)^{r/2}  &  0\\
\dfrac{1+(-1)^{r/2}}{2}  & 0 & \dfrac{1-(-1)^{r/2}}{2}\\                
\end{matrix}\right).
\end{equation}
There are two cases depending on the value of $r$.
If $r$ is odd multiple of two, then 
\begin{equation}
\label{}U^{2s} =
 \left( \begin{matrix}
1 & 0 & 0\\
0 & 1 & 0\\
0 & 0 & 1\\                
\end{matrix}\right)
\end{equation}
which is an identity matrix implying the periodicity of quantum correlations.
If $r$ is even multiple of two, then
\begin{equation}
\label{}U^{2s} =
 \left( \begin{matrix}
0 & 0   &  1\\
0 & -1  &  0\\
1 & 0 & 0\\                
\end{matrix}\right).
\end{equation}
Thus, $U^{2s}[a,b,c]^T=[c,-b,a]^T$. In the two-qubit basis 
$[c,-b,a]^T$ is equal to $c|1\rangle_1|1\rangle_2-(b/\sqrt{2})(|1\rangle_1|0\rangle_2+|0\rangle_1|1\rangle_2)+
a|0\rangle_1|0\rangle_2$. Again using the formula for concurrence for two-qubit pure state \cite{Wootters01} one
obtains $2|b^2/2-ac|$ for both the states, thus proving the claimed periodicity of quantum correlations.
It can be shown that the same results hold true for $p=3\pi/2$.

\subsection{Case of $p=\pi$}
For $p=\pi$ the Floquet operator reduces to
\begin{equation}
\label{}
\left(\begin{matrix}
0 & 0 &  e^{-ik/2}\\
0 & -1 & \\
e^{-ik/2} & 0 & 0\\                
\end{matrix}\right).
\end{equation}
Its eigenvalues and eigenvectors are respectively given as $\{ -e^{-ik/2} , e^{-ik/2} , -1 \}$,
$[-1/\sqrt{2},0,1/\sqrt{2}]^T$, $[1/\sqrt{2},0,1/\sqrt{2}]^T$ and $[0,1,0]^T$. Thus, using them the Floquet 
operator for $n$th time can be obtained and is given as follows:
\begin{equation}
\label{Eq:UNpi} U^n =
\frac{1}{2} \left( \begin{matrix}
\alpha & 0 & \beta\\
0 & (-1)^n & 0\\
\beta & 0 & \alpha\\                
\end{matrix}\right).
\end{equation}
where $\alpha=\left(-e^{-ik/2} \right)^n+\left(e^{-ik/2}\right)^n$ and 
$\beta=-\left(-e^{-ik/2} \right)^n+\left(e^{-ik/2}\right)^n$.
Consider the case of chaos parameter $k=r\pi/s$.
It will be proved that if $r$ is odd then the time period of quantum correlations is $T=2s$, otherwise, it is $T=s$.

{\it Odd $r$ :} 
In this case using Eq.~(\ref{Eq:UNpi}) one obtains:
\begin{equation}
\label{}
U^{2s}= \left( \begin{matrix}
-1 & 0 & 0\\
0 & 1 & 0\\
0 & 0 & -1\\                
\end{matrix}\right).
\end{equation}
It can be seen that $U^{2s}$ is a diagonal matrix and it is shown in an identical case in 
Sec.~\ref{sec:periodicity} that quantum correlations are invariant under its action. 
Apart from this periodicity of $2s$ additional temporal periodicity is also found.
For initial separable state the 
quantum correlations at times $t=s+l$ and $t=s-l$ are same for $1\leq l \leq s-1$. This argument can be extended 
to $t>2s$. Details of the derivation of this result are given in Appendix~\ref{Appendix1}.

{\it Even $r$ :}
Consider the case of even $r$ which implies odd $s$. It will be now shown that the period is $s$. 
Using Eq.~(\ref{Eq:UNpi}) one obtains:
\begin{equation}
\label{}
U^{s}= \left( \begin{matrix}
0 & 0 & i^{r}\\
0 & -1 & 0\\
i^{r} & 0 & 0\\                
\end{matrix}\right).
\end{equation}
Thus, if $r$ is odd multiple of $2$ then $U^{s}[a,b,c]^T=[-c,-b,-a]$ otherwise $U^{s}[a,b,c]^T=[c,-b,a]$.
It can be seen easily that the concurrence for both the state is $2|b^2/2-ac|$ proving the claimed
periodicity.

In this case, apart from this periodicity of $s$, additional temporal periodicity is found. For the initial 
separable state the quantum correlations at times $(s-2l-1)/2$ and $(s+2l+1)/2$ are same for $1\leq l \leq (s-3)/2$.
Details of the derivation of this result are given in Appendix~\ref{Appendix2}. It can be shown that the same 
results holds true for $p=0$ and $2\pi$.
It should be pointed here that no such time periodicity was observed for $j>1$ (as also shown in 
Fig.~\ref{fig:jis3by2Q1}, \ref{fig:jis3by2Q2}, \ref{fig:jis3Q1} and \ref{fig:jis3Q2}) even if $t>>1$.
It should also be pointed that these periodicities in $k$, and that of time for the case $j=1$, 
of quantum correlations are of purely quantum origin and are independent of the underlying classical phase space.

\section{Summary}

Quantum kicked top is a fundamental model of Hamiltonian chaos and has been
realized experimentally in various distinct test-beds, namely, the hyperfine
states of cold atoms, coupled superconducting qubits and recently in a two-qubit system using 
Nuclear Magnetic Resonance techniques \cite{krithika2018nmr}.
This model advantage that it can be represented in terms of qubits and
lends itself naturally to theoretical studies on the connections between
quantum correlation measures and classical dynamical properties.
With increasing interest in the experimental results using quantum kicked
top \cite{Shohini2018,ArulChaos2018}, this paper presents new results on the periodic behaviour of quantum
correlation measures (using $j$ spins to represent the kicked top) as a
function of either time or kick strength when certain conditions are satisfied.
Due to the periodicity of quantum correlations, experimentally it is sufficient
to explore the parameter space corresponding to the basic unit. This
work provides an upper bound on the parameter values corresponding to this
basic unit.

In particular, it is shown analytically as well as demonstrated numerically that, for a given initial quantum state,
the quantum correlations are periodic in kick strength $k$ with a period
given by $\kappa=2j\pi$. A special case of this result was reported in
Ref. \cite{Arjendu2017}. Since this is valid for large $j$, periodicity in $k$
is seen in the semiclassical limit as well. 
This has also been verified through numerical simulations
for bipartite measures of entanglement like the von Neumann entropy, quantum
discord and concurrence. Similar numerical results have also been obtained for
the multipartite entanglement measures such as 3-tangle and Meyer and Wallah $Q$
measure. The phase space of the kicked top for any given value of $k$ is
isomorphic to that at $-k$. This observation, when combined with the periodicity
of $\kappa=2j\pi$ shows that the unique signatures of phase space are
obtained only in the range $[0,j\pi]$. This can guide experimental implementations
of the kicked top on the appropriate choice of parameters, given the value of $j$.

Temporal periodicity of quantum correlations are analytically shown to arise
for $j=1$ (two qubit case) if $k=r\pi/s$, where $r$ and $s$ are integers if
the angular frequency $p$ can take any of the values from the set
$\{0,\pi/2,\pi,3\pi/2,2\pi\}$.
In the case of $p=\pi/2$, the period is shown to be $T=4s$ for odd $r$ otherwise it is $T=2s$, whereas for $p=\pi$
the period is shown to be $T=2s$ for odd $r$ otherwise it is $T=s$. In the case of $p=\pi$ (same results
hold true for $p=2\pi$) additional temporal periodicity are proved. If the initial state is separable then for 
odd $r$ it is shown that quantum correlations are same at $t=s+l$ and $t=s-l$ such that $1\leq l \leq s-1$. 
Whereas the same is true for even values of $r$ at times $(s-2l-1)/2$ and $(s+2l+1)/2$ such that
$1\leq l \leq (s-3)/2$. These results can be extended for times longer than the respective time
periods $T$.

The case of $j=1$ has one more experimental implication. Kicked top experiments are limited by
the coherence time  $\tau_{coh}$, which is typically not large. The entire experiment including
the read-out should be completed by this timescale.
If $k=r\pi/s$ and $p$ is chosen from the set $\{0, \pi/2,\pi, 3\pi/2,2\pi \}$,
then the period $T$ of quantum correlations as a function of time is known from the results
obtained in this work. Thus, the relevant time scale
for the experiments is $\mbox{min}(\tau_{coh},T)$. This implies that in some cases $T$ can be made smaller than
$\tau_{coh}$ effectively improving the reliability of the experimental results.

\section{Acknowledgments}
UTB gratefully acknowledges the discussions with T. S. Mahesh and V. R. Krithika.
UTB acknowledges the funding received from the Department of Science and Technology, India under the scheme 
Science and Engineering Research Board (SERB) National Post Doctoral Fellowship (NPDF) file number PDF/2015/00050.

\appendix
\section{Derivation of additional temporal periodicity for $p=\pi$ and odd $r$}
\label{Appendix1}

In this Appendix  additional temporal periodicity for $p=\pi$ and for odd $r$ in the value of $k=r\pi/s$ 
will be proved.
It will be proved that if the initial state $[a,b,c]^T$ is separable then the 
quantum correlations at time $t=s+l$ and $t=s-l$ are same for $1\leq l \leq s-1$.
We will restrict ourselves to 
time interval $[0,2s]$ and the argument can be extended to $t>2s$. Consider the case of odd $l$.
Then, $s\pm l$ will be odd. Thus, using Eq.~(\ref{Eq:UNpi}) one obtains
\begin{equation}
\label{}
U^{s\pm l}= \left( \begin{matrix}
0 & 0  & e^{-i r (s\pm l) \pi/2s}\\
0 & -1 & 0\\
e^{-i r (s \pm l) \pi/2s} & 0  & 0\\                
\end{matrix}\right).
\end{equation}
This implies
\begin{equation}
U^{s\pm l} [a,b,c]^T=[c\, e^{-i r (s\pm l) \pi/2s},-b,a\, e^{-i r (s\pm l) \pi/2s}].
\end{equation}
This can be written in the two-qubit basis as follows:
\begin{eqnarray}
\label{Eq:statesplusl}
&&c \,e^{-i r (s\pm l) \pi/2s}|1\rangle_1|1\rangle_2-(b/\sqrt{2})\left(|1\rangle_1|0\rangle_2+|0\rangle_1|1\rangle_2\right)\nonumber\\
&&\,\,\,\,\, +a \,e^{-i r (s\pm l) \pi/2s}|0\rangle_1|0\rangle_2.
\end{eqnarray}

Concurrences for $2-$qubit pure states in Eq.~(\ref{Eq:statesplusl}) are 
$2|b^2/2-a\,c\,e^{-i\, r\, (s\pm l)  \pi/s}|$. 
Since the initial state $[a,b,c]^T$ is separable the concurrence formula gives 
$ac=b^2/2$, the concurrence becomes
\begin{align}
 & 2|ac|~|1-e^{-ir(s\pm l)\pi/s}|  \nonumber \\
 & = 2|ac|\sqrt{2\left(1-\cos\left( r\pi \pm rl\pi/s \right) \right)}.
\end{align}
The cosines of both these angles are same since
they are reflection of each other about $x$-axis. Similarly, it can be shown for even $l$ that the quantum 
correlations at times $t=s+l$ and $t=s-l$ are same.

\section{Derivation of additional temporal periodicity for $p=\pi$ and even $r$}
\label{Appendix2}

In this Appendix  additional temporal periodicity for $p=\pi$ and for even $r$ in the value of $k=r\pi/s$
will be proved. It will be proved that if the initial state $[a,b,c]^T$ is separable then the quantum 
correlations at times $(s-2l-1)/2$ and $(s+2l+1)/2$ are same for $1\leq l \leq (s-3)/2$.
Consider the case of even $(s-2l-1)/2$ which implies $(s+2l+1)/2$ is odd since the difference between them 
is $2l+1$. Thus, using Eq.~(\ref{Eq:UNpi}) one obtains:
\begin{equation}
\label{}
U^{(s-2l-1)/2}= \left( \begin{matrix}
  e^{-i r (s-2l-1)\pi/(4s) } & 0 & 0\\
0 & 1 & 0\\
0 & 0 & e^{-i r (s-2l-1)\pi/(4s) } \\                
\end{matrix}\right)
\end{equation}
whereas 
\begin{equation}
\label{}
U^{(s+2l+1)/2}= \left( \begin{matrix}
0 & 0 & e^{-i r (s+2l+1)\pi/(4s) } \\
0 & -1 & 0\\
e^{-i r (s+2l+1)\pi/(4s) }  & 0 & 0\\                
\end{matrix}\right).
\end{equation}
This gives 
\begin{eqnarray}
\begin{split}
&U^{(s-2l-1)/2}[a,b,c]^T =\\
&[ a \,e^{-i r (s-2l-1)\pi/(4s) },b,c\, e^{-i r (s-2l-1)\pi/(4s) }]^T
\end{split}
\end{eqnarray}
while 
\begin{eqnarray}
\begin{split}
& U^{(s+2l+1)/2}[a,b,c]^T=\\
& [c \,e^{-i r (s+2l+1)\pi/(4s) } ,-b, a\, e^{-i r (s+2l+1)\pi/(4s) }]^T.
\end{split}
\end{eqnarray}
The concurrence for these states are then 
\begin{align}
&  2|b^2/2 -a\,c\, e^{-i r (s-2l-1)\pi/(2s)}|  ~~~ \mbox{and} \nonumber \\
& 2|b^2/2 -a\,c\, e^{-i r (s+2l+1)\pi/(2s)}|  \nonumber
\end{align}
respectively. Since the initial state $[a,b,c]^T$ is separable implies $ac=b^2/2$. Then the concurrences becomes
\begin{align}
&  2|ac|~|1 -e^{-i r (s-2l-1)\pi/(2s)}| ~~~ \mbox{and} \nonumber \\
&  2|ac|~|1 -e^{-i r (s+2l+1)\pi/(2s)}|  \nonumber
\end{align}
respectively which can be written as
\begin{align}
&  2\sqrt{2}|ac|~|1 -\cos\left( r (s-2l-1)\pi/(2s) \right)| ~~~ \mbox{and} \nonumber \\
&   2\sqrt{2}|ac|~|1 -\cos\left( r (s+2l+1)\pi/(2s) \right)| \nonumber
\end{align}
respectively. It can be seen that for $r$ even $\cos\left( r (s-2l-1)\pi/(2s) \right)$ and 
$\cos\left( r (s+2l+1)\pi/(2s) \right)$ are equal since the angles are reflection of each other about $x$-axis. 
Similarly, this result can be proved for odd $(s-2l-1)/2$.
\bibliography{reference22013,reference22}
\end{document}